**Title**
Tianwen-2 target asteroid (469219) Kamoʻoalewa probably develops an Itokawa-compositional but ultra-highly space-weathered surface

**Authors**
Pengfei Zhang[1], Guozheng Zhang[2], Zichen Wei[3], Mikael Granvik[4,5], Xiaoran Yan[6], Pengyue Wang[2], Qinwei Zhang[7], Ronghua Pang[1], Wen-Han Zhou[8], Te Jiang[3], Pierre Vernazza[9], Takahiro Hiroi[10], Edward Cloutis[11], Francesca DeMeo[12], Pierre Beck[13], Wing-Huen Ip[14], Marco Fenucci[15,16], Yongxiong Zhang[17], Michael Marsset[18], Yunbo Niu[19], Xuejin Lu[20], Xing Wu[21], Honglei Lin[22], Shoucun Hu[7], Bin Cheng[23], Haibin Zhao[7], Xiaobin Wang[24], Xiaoping Lu[25], Yonglong Zhang[26], Zongcheng Ling[20], Jiang Zhang[20], Sizhe Zhao[2], Cateline Lantz[27], Jooyeon Geem[28], Zhiping He[29], Juntao Wang[30], Liyong Zhou[31], Xiliang Zhang[24], Shijei Li[1], Sen Hu[22], Wei Yang[22], Xiongyao Li[1], Xiaoping Zhang[2], Jiahui Liu[2], Peng Zhang[32], Guang Zhang[32], Yangting Lin[22], Yang Li[1]*

**Affiliations**
[1]Center for Lunar and Planetary Sciences, Institute of Geochemistry, Chinese Academy of Sciences, Guiyang, China.
[2]State Key Laboratory of Lunar and Planetary Sciences, Macau University of Science and Technology, Macau, China.
[3]School of Earth Sciences, China University of Geosciences, Wuhan, China.
[4]Department of Physics, University of Helsinki, Helsinki, Finland.
[5]Asteroid Engineering Laboratory, Luleå University of Technology, Kiruna, Sweden.
[6]Institute of applied physics "Nello Carrara", Sesto Fiorentino, Italy.
[7]Key Laboratory of Planetary Sciences, Purple Mountain Observatory, Chinese Academy of Sciences, Nanjing, China.
[8]Department of Earth and Planetary Science, The University of Tokyo, Tokyo, Japan.
[9]Aix Marseille Université, CNRS, CNES, Laboratoire d'Astrophysique de Marseille, Marseille, France.
[10]Department of Geological Sciences, Brown University, Providence, USA.
[11]Department of Geography, University of Winnipeg, Winnipeg, Canada.
[12]Department of Earth, Atmospheric, and Planetary Sciences, Massachusetts Institute of Technology, Cambridge, USA.
[13]UJF-Grenoble 1/CNRS-INSU, Institut de Planétologie et d'Astrophysique de Grenoble, Grenoble, France.
[14]Institute of Astronomy, National Central University, Chung Li, Taiwan.
[15]ESA ESRIN/PDO/NEO Coordination Centre, Largo Galileo Galilei, Frascati, Italy.
[16]Elecnor Deimos, Via Giuseppe Verdi, San Pietro Mosezzo, Italy.
[17]School of Engineering, Guangzhou College of Technology and Business, Guangzhou, China.
[18]European Southern Observatory (ESO), Karl-Schwarzschild-Strasse 2, 85748 Garching bei München, Germany.
[19]Chongqing University State Key Laboratory of Power Transmission Equipment and System Security and New Technology, Institute of Systems Engineering, Chongqing, China.
[20]Shandong Key Laboratory of Optical Astronomy and Solar-Terrestrial Environment, School of Space Science and Physics, Institute of Space Sciences, Shandong University, Weihai, China.

[21]State Key Laboratory of Space Weather, National Space Science Center, Chinese Academy of Sciences, Beijing, China.
[22]Key Laboratory of the Earth and Planetary Physics, Institute of Geology and Geophysics, Chinese Academy of Sciences, Beijing, China.
[23]School of Aerospace Engineering, Tsinghua University, Beijing, China.
[24]Yunnan Observatories, Chinese Academy of Sciences, Kunming, China.
[25]School of Computer Science and Engineering, Faculty of Innovation Engineering, Macau University of Science and Technology, Macau, China.
[26]The College of Mechanical and Vehicle Engineering, Taiyuan University of Technology, Taiyuan, China.
[27]Institut d'Astrophysique Spatiale, CNRS/Université Paris Saclay, Orsay, France.
[28]Department of Physics and Astronomy, Seoul National University, Seoul, Republic of Korea.
[29]Shanghai Institute of Technical Physics, Chinese Academy of Sciences, Shanghai, China.
[30]Institute of Deep Space Sciences, Deep Space Exploration Laboratory, Hefei, China.
[31]School of Astronomy and Space Science & MOE Key Laboratory of Modern Astronomy and Astrophysics, Nanjing University, Nanjing, China.
[32]Technology and Engineering Center for Space Utilization, Chinese Academy of Sciences, Beijing, China.

**\*Corresponding Author.** E-mail: liyang@mail.gyig.ac.cn

**Abstract**

China's Tianwen-2 mission plans to return samples from a small, rapidly spinning Earth quasi-satellite (469219) Kamoʻoalewa. Previous studies linked Kamoʻoalewa to lunar composition and origin. Here, we propose another scenario. We reanalyzed the reflectance spectrum of Kamoʻoalewa and obtained an absorption band center at 1.001±0.028 μm (error is 1σ), consistent with LL chondrites. We then conducted space weathering (SW) experiments on meteorites and found that highly space-weathered LL chondrite powder (but not slab) successfully reproduced the reflectance spectrum of Kamoʻoalewa. We further traced the dynamical origin of Kamoʻoalewa and found that it probably originated from the $\nu_6$ secular resonance, and more specifically, the Flora family. Kamoʻoalewa exhibits a similar composition to Itokawa and 7 objects in the Flora family, but with a higher degree of space weathering. We, therefore, proposed that Kamoʻoalewa probably originated from the Flora family and developed an Itokawa-compositional, highly space-weathered, fine-regolith-dominated surface.

## Introduction

Near-Earth asteroids (NEAs) are small objects orbiting in near-Earth space. Over the past two decades, two questions have received particular attention: (1) What are the sources of NEAs and terrestrial meteorites? (2) How do the surfaces of NEAs evolve? Thanks to ground-based spectroscopic surveys, the composition distribution of NEAs is now increasingly clearly mapped[1–3]. In particular, Q-, Sq-, and S-type asteroids are the parent bodies of ordinary chondrites (OCs)[4,5] and account for approximately 65% of the number[6] and more than half of the mass[1] of NEAs. Of these, about two-thirds are spectrally linked to LL chondrites[7], a low-iron and low-metal group that accounts for only 10% of terrestrial OCs (47% for L and 43% for H). This difference in frequency distribution has been discussed as being related to the sizes, source regions, and impact ages of the NEAs[1,7]. For example, kilometer-sized NEAs with LL chondrite composition predominantly migrated through the $\nu_6$ secular resonance and originated from the old (1.2±0.2 billion years[8], yr) Flora family[1,9,10]. The meter-sized population is the main contributor to the Earth's LL group falls and has a wider source region due to its stronger susceptibility to the Yarkovsky effect[1,7]. In addition, older $^{40}Ar/^{39}Ar$ main belt impact ages of the LL group[11] suggest that LL chondrite-like fragments migrated to near-Earth space and fell to Earth earlier than L and H chondrites[1]. This is consistent with recent studies that young asteroid families, the Karin family (~5.8 × 10$^6$ yr) and the Koronis$_2$ family (~7.6 × 10$^6$ yr), are the primary sources of H chondrites[8], while the Massalia$_2$ family (~4 × 10$^7$ yr) contributes mainly to L chondrites[12]. Nevertheless, material returned from these meteorite parent bodies could provide the most direct evidence of these relationships.

In 2010, the Hayabusa spacecraft successfully returned regolith grains from the Sq-type NEA (25143) Itokawa[13]. Before the launch of the probe, Itokawa was linked to be an LL chondrite composition through ground-based observations[14]. Subsequent petrologic[15] and oxygen isotopic analysis[16] of the returned grains linked Itokawa to the LL5/6 chondrites. Meanwhile, detailed orbital[17,18] and laboratory[19–21] investigations of space weathering (SW) characteristics provided insights into the evolution of Itokawa. SW refers to the process by which the surface features of airless objects are altered by micrometeoroid bombardment and solar wind[22]. For the Moon and asteroids with OC composition, SW produces sub-micro-sized iron (npFe$^0$) particles, reduces reflectance (darkening), decreases absorption band depths, and increases spectral slope (reddening)[23]. SW makes the asteroid surface mature (older) and drives the evolution of Q- (younger surface) to Sq- (middle-aged surface) and S-type (older surface)[24]. In contrast, resurfacing processes[25] create a fresher (younger) surface and cause S- to revert to Sq- and Q-type[1]. Itokawa underwent an intermediate timescale (10$^3$ to 10$^6$ yr) of SW reddening[26–29] and experienced non-global localized resurfacing[29,30]. Since Itokawa is an intermediate object between SW-fresh (Q-type) and SW-strong (S-type)[1], it serves as a key reference for studying the relationship between the spectral evolution of silicate-rich asteroids and the SW timescale. Q-type asteroids are the representatives of SW-fresh and have been widely discovered[1,3]. However, it is still unclear whether OC-like asteroids can evolve into SW-strong surfaces.

Now, China's Tianwen-2 mission promises to shed further light on the origin and evolution of NEAs. The Tianwen-2 probe was launched on 29 May 2025 and plans to return samples from NEA (469219) Kamoʻoalewa in 2027 and then orbit the main belt active asteroid 311P/PANSTARRS in 2034[31]. Kamoʻoalewa is a quasi-satellite of the Earth[32] that is less than 100 meters in size and has a short rotation period of approximately 28 minutes[33]. Previous studies[33,34] noticed that Kamoʻoalewa displays an extremely red spectral slope and silicate-like absorption feature near 1 μm. Based on spectral properties and orbital dynamical analyses, Kamoʻoalewa was previously implicated as being of lunar origin[33–36]. However, previous studies[33,34,36] did not carefully analyze absorption characteristics or consider the asteroid's SW effect. Here, we conducted new analyses on spectroscopy, SW experiments, and orbital dynamics. Our results suggest that Kamoʻoalewa is probably an Itokawa-compositional, highly space-weathered object originating from the $\nu_6$ secular resonance, more specifically, the Flora family.

## Results

**Reflectance spectrum and composition of Kamoʻoalewa**

We first analyzed the visible to near-infrared (VIS-NIR) reflectance spectrum of Kamoʻoalewa, previously obtained by the Large Binocular Telescope (LBT)[33]. As previously reported[33], Kamoʻoalewa shows an obvious absorption around 1 µm (hereafter Band I, Fig. 1a). Band I absorption is generally contributed by crystal field transitions of ferrous iron ($Fe^{2+}$) in olivine or pyroxene[37,38]. It is mainly related to the $Fe^{2+}$ content, and different types of meteorites show different Band I centers and Band I depths[39]. Simulation experiments on olivine[40–42] and $Fe^{2+}$-bearing silicate meteorites[43,44] have suggested that SW does not generally shift the Band I center more than ±10 nm. Band I center is, therefore, diagnostic and has been widely used to determine the composition of meteorites[39] and asteroids such as Itokawa[15]. Kamoʻoalewa's Band I center is also calculated here.

In the first step, we renormalized Kamoʻoalewa's observed spectra at 0.55 µm (see Fig. 1a). In the second step, due to the low signal-to-noise ratio of the observed spectra (with an error of 1σ[33]), we generated 10,000 spectral curves using the Monte Carlo method. In the third step, we determined the two endpoints of a straight-line continuum on either side of the Band I absorption feature. The generated 10,000 spectra exhibit various curve shapes (some curves oscillate repeatedly, unlike any type of asteroid), and using the traditional method[45] to determine the continuum for each spectrum may lead to misidentification of the Band I center. Therefore, for each spectral curve, we fixed the wavelengths corresponding to the left and right endpoints of the continuum at two wavelengths. These two wavelengths were determined from the main spectrum (composed of black circular data points in Fig. 1a) because it is the most reliable compared to other generated spectra. For the main spectrum, on the right side of Band I, the data point at 1.230 µm (normalized reflectance is 1.534) is close to the data point at 1.247 µm (normalized reflectance is 1.484), while the data point at 1.253 µm (normalized reflectance is 1.613) is far from the date point at 1.247 µm (normalized reflectance is 1.484). Therefore, the point at 1.230 µm was considered to have high measurement reliability and was selected as the right endpoint of the continuum. Since the data points of the main spectrum are discrete on the left side of Band I, 2nd to 3rd-order polynomial fits were performed for data points within 0.6–0.85 µm. The data point at 0.664 µm (normalized reflectance is 1.095) is closest to the average of the tangent points from fitted curves and is therefore selected as the left endpoint of the continuum (Supplementary Fig. 1). In this way, 0.664 µm and 1.230 µm were fixed as wavelength sites of the left and right endpoints of the continuum. In the fourth step, we divided out the continuum for each generated spectrum. In the fifth step, for each continuum-removed spectrum, we performed 3rd-order polynomial fit for data within 0.85–1.15 µm. Past experience in analyzing spectra of meteorites and asteroids suggested that this fitting method effectively detects the Band I centers and depths[46]. Note that we only retained the results where the fitted curve was concave, totaling 4637 spectra. Thereby, we obtained a Band I center of 1.001±0.028 µm (error is 1σ), and a corresponding band depth of 15.66±2.14% (error is 1σ), as shown in Fig. 1b. This Band I center range largely overlaps with LL chondrite and partially overlaps with L chondrite and andesitic achondrites (Fig. 1b), implying that Kamoʻoalewa probably has an LL chondrite, L chondrite, or andesitic achondrite composition. Since the low signal-to-noise ratio and limited wavelength range do not allow for the calculation of Kamoʻoalewa's band area ratio, it is impossible to determine from Fig. 1b alone which of LL, L, or andesitic achondrite corresponds to Kamoʻoalewa. However, other spectral parameters, such as spectral slope and Band I depth, are also related to composition and could assist in further determining Kamoʻoalewa's composition.

As previously noted[33,34], Kamoʻoalewa shows an extremely red VIS-NIR spectral slope ($0.723^{+0.076}_{-0.071}$ µm$^{-1}$ here, calculated by least squares linear fit in the range 0.45–2.194 µm). Generally, redder spectral slopes on silicate-rich asteroid surfaces can result from smaller grain sizes or higher porosity[47], larger phase angles[48], higher metal content[49,50], and SW processes[22]. However, laboratory measurements for LL/L chondrites[47] and andesitic achondrite[51] suggest that even the finest fresh powders cannot account for the Kamoʻoalewa-like red slope. The silicate slabs that present the lowest porosity generally show a bluer spectral slope than powders that present the highest porosity[44,47,51]. Laboratory measurements for LL and L chondrites indicate that even the largest phase angle variations (13°–120°) cannot generate spectral slopes as red as Kamoʻoalewa[48], which has an observed phase angle of 43.2°[33]. While there is a current lack of photometric studies on andesitic meteorites, a spectrum measured at 30° phase angle for an andesitic achondrite is significantly bluer than that of Kamoʻoalewa[51]. Additionally, spectral calculations for silicate-metal mixtures indicate that achieving both the Kamoʻoalewa-like red spectral slope and absorption feature is impossible[33,50]. Some metal-bearing meteorites, such as enstatite chondrites, pallasites, mesosiderites, and iron meteorites, lack the 1.001±0.028 µm absorption (Fig.1b and Methods), even though they exhibit red spectral slopes[49,52]. Next, we focus on testing whether the SW of LL/L chondrites and andesitic achondrite can produce the Kamoʻoalewa-like red spectral slope.

We first used a nanosecond-pulsed laser to irradiate an LL5/6 chondrite, Kheneg Ljouâd, simulating the SW spectral alteration effects caused by micrometeoroid bombardments on asteroids composed of LL chondrite. Different irradiation energies were set to simulate various degrees of SW (Methods). A slab and a loose powder sample (grain size < 45 µm) were irradiated to respectively represent two typical endmembers of the SW pattern: (1) a surface composed solely of boulders, rocks, pebbles, or coarse grains, and (2) a surface covered with fine regolith particles or dust. As shown in Fig. 2, even if the simulated SW is saturated for the slab, the corresponding spectral slope remains much flatter than Kamoʻoalewa's. Currently, no experiments suggested that LL slabs can be space weathered to produce a Kamoʻoalewa-like red spectral slope. Recently, a thermal inertia derived from the newest Yarkovsky effect observation for Kamoʻoalewa gives a very low value, indicating that it may develop a fine-grained regolith with an average size of 100 µm to 3 mm[53]. Three simulations works[54–56] also suggested that fine-grained regolith can be retained on the surface of a Kamoʻoalewa-like small and fast-rotating asteroid. Therefore, the first SW pattern (for slabs) was not considered further in this study. However, using a 40 mJ × 80 times energy to irradiate the powder sample, which simulated the SW timescale of $9.33 \times 10^6$ yr at 1 AU (Methods), we obtained a reflectance spectrum closely matching that of Kamoʻoalewa. After irradiation, the powder sample darkens significantly (Fig. 3a–b), and the visible reflectance at 0.55 µm decreases sharply from an initial value of 0.303 to 0.085 (Fig. 3c, Supplementary Fig. 2a). After removing the spectral continuum and performing the 3rd-order polynomial fit, the Band I center of the irradiated powder shifts from 1.000 µm to 0.985 µm, and the corresponding band depth decreases from 30.76% to 14.85%. This is very close to Kamoʻoalewa's Band I center of 1.001±0.028 µm and Band I depth of 15.66±2.14%. In particular, the irradiated powder exhibits a red spectral slope similar to Kamoʻoalewa (Fig. 3d, Supplementary Fig. 2b). Further, our microscopic observations suggested that abundant agglutinates, amorphous rims, and npFe$^0$ particles, which are the main products of asteroid SW processes[57], appear in the irradiated powder (Fig. 4).

We then conducted laser irradiation experiments on an L chondrite powder sample (grain size < 45 µm). As a result, neither fresh nor irradiated L chondrite powders produced Kamoʻoalewa-like red spectra (Fig. 5a–b). We also irradiated an H chondrite powder (size < 45 µm) and found that both fresh and weathered powders could not produce spectra matching that of Kamoʻoalewa (Fig. 5c–d). These results are not surprising, because compared to LL chondrite, L and H chondrites generally have higher pyroxene content, lower olivine content, and lower $Fe^{2+}$ content in olivine and pyroxene[39,58]. Olivine is more susceptible to space weathering than pyroxene[40,59], and a higher $Fe^{2+}$ content facilitates the generation of $npFe^0$[60], leading to easier reddening. Currently, no other experiments have been reported that L and H chondrites can be space weathered to a Kamoʻoalewa-like red slope. Although we did not irradiate the L and H slabs here, considering that slabs generally have a bluer slope than powders[44,47] and the irradiated result of LL (even if the slab is saturated with space weathering, it cannot produce a Kamoʻoalewa-like red spectrum, but irradiating the powder can), L and H slabs may be more difficult to produce spectra as red as those of Kamoʻoalewa. Additionally, although we did not irradiate the andesitic meteorite, given that it is mainly composed of pyroxene, feldspar, and quartz[51], and that its spectrum is similar to HED meteorites[47,51] (daughter of V-type asteroids, rich in pyroxene[58] and extremely insensitive to space weathering[61]), it seems less likely to contribute a Kamoʻoalewa-like red spectrum than the H and L chondrites.

Previous studies[33,34] found that only three lunar materials (Apollo 14 soil, Lunar 24 soil, and Yamato,791197,72 meteorite) display both extremely red spectral slopes and Band I absorption features like Kamoʻoalewa (Fig. 6a); however, their Band I center and Band I depth were not calculated and compared with those of Kamoʻoalewa. Here, using the same method as the LL chondrite, we calculated the Band I centers and Band I depths for these three lunar materials. None of them, but only laser-irradiated (at 40 mJ × 80 times energy) LL chondrite powder, falls within the 1σ range of Kamoʻoalewa's Band I center and Band I depth (Fig. 6b). The Apollo 14 soil and Yamato,791197,72 meteorite are not within the 1σ range of Kamoʻoalewa's Band I center (Fig. 6b). Given that Band I center is diagnostic of composition, this suggests that Kamoʻoalewa is not similar to Apollo 14 soil and Yamato,791197,72 in composition. Although Lunar 24 soil falls into the 1σ range of Kamoʻoalewa's Band I center (Fig. 6b), it shows a redder slope than Kamoʻoalewa. If Kamoʻoalewa has the same composition as Lunar 24 soil, the redder slope means a higher degree of space weathering, then Lunar 24 soil should show a shallower Band I depth than Kamoʻoalewa. However, as shown in Fig. 6b, this is not the case. Our irradiated LL chondrite powder falls into the 1σ range of Kamoʻoalewa's Band I center, indicating a similarity in the composition. Meanwhile, it shows a slightly redder spectral slope than Kamoʻoalewa, meaning that we simulated a slightly higher degree of space weathering than Kamoʻoalewa. If Kamoʻoalewa is indeed similar to LL chondrite in composition, then Kamoʻoalewa should show a slightly deeper Band I depth than laser-irradiated LL chondrite powder. As shown in Fig. 6b, this is indeed the case. These indicate that the space-weathered LL chondrite is more credible than these three lunar materials. However, it should be specifically stated that although these three lunar materials do not match Kamoʻoalewa better than laser-irradiated LL chondrite powder in Fig. 6b, given the lunar diverse composition and the current low spectral signal-to-noise ratio of Kamoʻoalewa, we cannot conclude here that lunar composition has been completely ruled out. For example, a recent study[36] investigated the spectra of many areas on the lunar surface and found that some of them are similar to the spectral curve of Kamoʻoalewa. We just emphasize that here, based on the results above, from the perspective of laboratory spectra alone, LL chondrite-compositional, highly space-weathered surface with fine regolith particles or dust for Kamoʻoalewa, is also a likely scenario and should be considered. The debate regarding the lunar and LL chondrite composition of Kamoʻoalewa will be discussed further in a later section.

Notably, as simulation experiments for such a high degree of SW on meteorites were previously lacking, our irradiation experiment on LL chondrite provides the first laboratory evidence that silicate-rich objects can develop extremely red surfaces through SW processes. Nevertheless, given the differences between laboratory simulations and asteroid environments, a subsequent question arises: Does such strong SW occur on asteroid surfaces? If so, there may be multiple silicate-rich asteroids as red as Kamoʻoalewa.

**Investigation of other extremely red asteroids**

To achieve this objective, we investigated the spectral slope of all NEAs and main belt asteroids (MBAs) reported in the MITHNEOS[1] and SMASS II[62] projects. As shown in Fig. 7a, among silicate-rich NEAs, A-type (1951) Lick and Srw-type (4142) Dersu-Uzala exhibit higher slope (1.205 μm$^{-1}$ and 0.851 μm$^{-1}$, respectively) compared to Kamoʻoalewa ($0.723^{+0.076}_{-0.071}$ μm$^{-1}$). Among silicate-rich MBAs, A-type (863) Benkoela shows a higher slope (1.028 μm$^{-1}$) than Kamoʻoalewa (Fig. 7b). A-type asteroids have been considered extremely rich in olivine[63]. Olivine is more susceptible to SW than pyroxene[40,59] and its spectral slope is more prone to reddening. Sr-type objects are also linked to LL chondrites[64]. Considering further that our SW experiment on LL chondrite powder produced extremely red spectra, it is likely that asteroids Lick, Dersu-Uzala, and Benkoela have undergone a high degree of SW.

Note that several D-type asteroids (D-type is usually featureless, and even if there are absorption features, the band depth does not exceed 3%) also show redder slopes than Kamoʻoalewa, which may be attributed to their presumed organics-rich composition. For example, Tagish Lake, a carbonaceous meteorite proposed as the analog of D-type, exhibits an extremely red spectral slope[65]. The redder slopes on D-type asteroids may also result from submicron-sized anhydrous interplanetary dust particles, which were proposed as analogs for D- and P-type objects[66]. The SW effect may not explain why D-type asteroids show such red slopes. After SW, a D-type asteroid (596) Scheila[67] and meteorite Tagish Lake[68,69] become more blue-sloped rather than red-sloped, although reddening versus bluing of organics-rich compositions is not yet fully resolved[70]. Therefore, although only four extremely red, silicate-rich asteroids were identified here, our investigation suggests that highly space-weathered objects may not be as scarce as previously thought.

**Dynamical source region and composition of Kamoʻoalewa**

An interesting question is, if Kamoʻoalewa is not a lunar but an LL chondrite compositional object, where does it originate? As an attempt to answer this question, here, we first use the NEA population seven-escape-region evolution model[10] combined with Kamoʻoalewa's current orbit and absolute magnitude, to trace the dynamical source region of Kamoʻoalewa (Methods). This procedure is composition-independent and can be cross-validated with spectrally identified mineralogy. As a result, Kamoʻoalewa shows a probability of 72±5% originating from the ν$_6$ secular resonance, 20±5% from the Hungaria region, and 8±1% from the 3:1 mean-motion resonance with Jupiter. This result aligns with a previous study on the orbital dynamics of Kamoʻoalewa[71]. The ν$_6$ secular resonance is adjacent to the Flora family, the major source region of NEAs with LL chondrite composition[1,7,12,72]. The Hungaria region contains a background population and the Hungaria family, where the background population is mainly composed of silicate-rich asteroids with L chondrite and primitive achondrite compositions[73]. The Hungaria family is mainly composed of Tholen E-type asteroids, the parent bodies of enstatite achondrites[74]. E-type asteroids lack the 1.001±0.028 μm absorption feature (most are featureless, and some have weak absorption at 0.49, 0.89–0.9, or 1.8 μm) and show a reddish NIR spectral slope[75]. The 3:1 resonance primarily sends H and L chondrite-like objects into near-Earth space[1]. Such a high probability (72±5%) suggests that, if Kamo'oalewa is not a lunar composition, it most likely originated from the $\nu_6$ secular resonance.

The sub-hundred-meter size of Kamo'oalewa[33] also allows us to use the newest orbital dynamic model, METEODOM[8], to trace its source region. This model contains 52 asteroid families as source regions and requires presetting the composition of a small object. Since the above spectroscopic analyses for Kamo'oalewa mainly point to an LL chondrite composition, we input the composition in the model as LL chondrites. As a result, Kamo'oalewa shows a probability of 81.7% originating from the Flora family and 18.3% from the Nysa family. The Nysa family is composed of E-type asteroids and LL-chondrite-composition objects[12]. Such a high probability (81.7%) suggests that, if Kamo'oalewa is an LL chondrite compositional object, the Flora family is its most likely source region.

Notably, while early dynamics studies focused on the scenario that Kamo'oalewa originated from the Moon and confirmed its orbital feasibility, they did not consider a main-belt source[33–36,76]. Conversely, our models do not include the Moon as a source region. Therefore, our results are not intended to refute the lunar origin of Kamo'oalewa, nor to compare the probability of lunar origin with main-belt origin. Rather, we demonstrate here that if Kamo'oalewa is not a lunar but an LL chondrite compositional object, then the $\nu_6$ secular resonance, and further, the Flora family, is the most likely source region. Recently, a quantitative estimate[77] showed that population models of NEAs based on the migration of objects from the main belt can explain the existence of Kamo'oalewa-like objects, without relying on the lunar origin hypothesis. Furthermore, numerical simulations[77] indicate that the expected number of Kamo'oalewa-like objects originating from the main belt is more than one order of magnitude larger than the expected number of those originating from the Giordano Bruno crater (this crater was previously believed to be Kamo'oalewa's most likely ejection site on the Moon[34]). In addition, recent numerical simulations[78] found pathways from the Flora family, and from the region near the 3:1 mean-motion resonance with Jupiter, to the Earth co-orbital region. These findings indicate that the dynamic source region of Earth's quasi-satellites may not be limited to the Moon. Regarding the lunar and main belt origins of Kamo'oalewa, we will continue to discuss in a later section.

**Comparison of the spectra of Kamo'oalewa, Itokawa, and Flora family**
Due to our above results and analyses pointing to an LL chondrite composition for Kamo'oalewa, here, we compare the reflectance spectra of Kamo'oalewa and LL chondrite-like NEA Itokawa. Kamo'oalewa exhibits a redder spectral curve (Fig. 8a) and a shallower Band I depth (Fig. 8b) than Itokawa, consistent with Kamo'oalewa being more space-weathered than Itokawa. We also compare the reflectance spectra of Kamo'oalewa and asteroid (43) Ariadne, an LL chondrite-compositional object with the second largest size in the Flora family[12,79]. Kamo'oalewa exhibits a redder spectral curve (Fig. 8a) and a slightly shallower Band I depth (Fig. 8b) than Ariadne, indicating that Kamo'oalewa is more space-weathered than Ariadne. Note that the Band I centers of Itokawa and Ariadne also fall within the 1σ range of Kamo'oalewa.

We also compare the spectra of Kamo'oalewa and the other 24 objects in the Flora family (these spectral data are from early observations[12]). Besides Itokawa and Ariadne, 6 objects (352 Gisela, 453 Tea, 913 Otila, 1667 Pels, 1857 Parchomenko, and 2873 Binzel) fall within the 1σ range of Kamo'oalewa's Band I center and show deeper Band I depths than Kamo'oalewa's average value (Fig. 8b, note that only 11 Flora family objects have complete VIS-NIR spectra that can be used for band parametric calculations). Additionally, all 25 bodies in the Flora family exhibit a bluer slope than

Kamoʻoalewa (Fig. 8a). These indicate that 7 objects in the Flora family (43 Ariadne, 352 Gisela, 453 Tea, 913 Otila, 1667 Pels, 1857 Parchomenko, and 2873 Binzel) are all highly similar to Kamoʻoalewa in composition but have a lower degree of space weathering than Kamoʻoalewa. This is unsurprising as the SW rate of NEAs is about 10 times faster than that of the MBAs[80] due to stronger solar wind radiation and lower-velocity micrometeoroid impacts. However, note that MBAs generally exhibit a slightly redder spectral slope than NEAs[81,82] because NEAs typically have a shorter collision lifetime and a stronger resurfacing[1,25]. However, note that the rule that MBAs are generally redder than NEAs applies to large objects (because they can be observed), and may not apply to small, fast-rotating Kamoʻoalewa-like objects. In a later section, we attempt to discuss why Kamoʻoalewa exhibits such a red spectrum.

## Discussion

### Composition and origin of Kamoʻoalewa

The previous two studies[33,34] linked Kamoʻoalewa to lunar composition and origin for several reasons: (1) no other extremely red-sloped asteroids were found besides Kamoʻoalewa, (2) only three lunar materials (Fig. 6a) display both extremely red spectral slopes and Band I absorption features like Kamoʻoalewa, and (3) simulations of orbital dynamics gave the feasibility of Kamoʻoalewa originating from the Moon. Further, the dynamical calculation indicated that if Kamoʻoalewa originated from the Moon, the young lunar crater "Giordano Bruno" is the most probable source region[34]. However, the following three facts should be considered: (1) As shown in Fig. 7, Kamoʻoalewa isn't the only extremely red asteroid rich in silicates. When early studies[33,34] conducted a similar spectral slope investigation to that in Fig. 7, the database of asteroid spectra used was too small to find other extremely red, silicate-rich asteroids. (2) The extremely red spectral slope is not unique to lunar material. Our laser irradiation experiment, shown in Fig. 3, has suggested that highly space-weathered LL chondrite powder can also produce a red spectrum like Kamoʻoalewa. The earlier two studies[33,34] did not conduct space weathering experiments, resulting in a lack of detailed consideration of this scenario. (3) Although the dynamical simulations[34] seem plausible, spectroscopy does not support a connection between Kamoʻoalewa and the "Giordano Bruno" crater. As shown in Fig. 8a, the "Giordano Bruno" crater shows a red spectral slope and weak Band I absorption, indicating that it has experienced a certain degree of SW. If Kamoʻoalewa originated from the "Giordano Bruno" crater, the deeper Band I absorption of the former than the latter means a higher degree of SW on Kamoʻoalewa, then the spectral slope of Kamoʻoalewa should be less red than the "Giordano Bruno" crater, but that's not the case (Fig. 8b). In addition, the Band I center of "Giordano Bruno" crater does not match that of Kamoʻoalewa (Fig. 8b). The "Giordano Bruno" crater, therefore, is implausibly the source region of Kamoʻoalewa. If, as revealed by early dynamical simulation[34] that "Giordano Bruno" crater is the most likely source region of Kamoʻoalewa on the Moon, then Kamoʻoalewa shouldn't originate from the Moon.

Our analyses of Band I centers, Band I depths, and spectral slope above suggested that the laser-irradiated LL chondrite powder matches Kamoʻoalewa, while three lunar materials (Apollo 14 soil, Lunar 24 soil, and Yamato,791197,72 meteorite) do not (Fig. 6). If as stated in two early studies[33,34], these three lunar materials are the samples in the laboratory spectral database that best match Kamoʻoalewa's spectrum (excluding our irradiated LL chondrite powder), then, the spectral evidence used to support their view of lunar origin appears tenuous. This at least means that our space-weathered

LL chondrite powder is more credible than these three lunar materials as an analogue of Kamoʻoalewa's surface. Additionally, early dynamical simulations supporting the lunar origin view assumed that Kamoʻoalewa originated from the Moon and verified it could have evolved into its current orbit, but did not consider a main-belt origin[33–36,76]. Although our two models used in this study did not include the Moon, recent two dynamical simulations[77,78] have revealed that the main belt also serves as a source region for Kamoʻoalewa-like Earth quasi-satellites. Further, the simulation[77] gave a comparative probability for Kamoʻoalewa to originate from the main belt of the Giordano Bruno crates, suggesting that the former is extremely unlikely. This study[77] also reproduced and predicted the emergence of other Earth quasi-satellites at size smaller than Kamoʻoalewa, indicating that the main belt can explain the results observed today, without taking into account the Moon as a source region. These challenge the early popular view that the Moon is the only dynamic source region of Kamoʻoalewa.

However, these findings do not mean our results completely denied the view of Kamoʻoalewa's lunar composition because (1) we did not investigate the spectra of all lunar materials and the lunar surface, which is not the focus of this study; and (2) the spectral slope is more sensitive to space weathering than Band I center and Band I depth, and it is not a key diagnostic parameter for identifying composition, serving merely as an auxiliary parameter. Given the diverse composition of the Moon, currently, we cannot completely rule out the possibility that a lunar material, or a lunar surface site, exhibits a Band I center within the 1σ range of Kamoʻoalewa. If a material or surface site matches the Band I center of Kamoʻoalewa but exhibits a redder spectral slope with shallower Band I depth or a bluer spectral slope with deeper Band I depth than Kamoʻoalewa, then theoretically, it is possible that it has a similar composition to Kamoʻoalewa, only differing in the degree of space weathering, just like our irradiated LL chondrite powder (Fig. 6) and spectra of Itokawa and 6 Flora family objects (Fig. 8). Perhaps, some lunar materials and lunar surface sites will also be like this. Considering the low signal-to-noise ratio of Kamoʻoalewa's spectrum again, therefore, we emphasize that we support the view that Kamoʻoalewa's surface is dominated by highly space-weathered LL chondrite-compositional fine-grained regolith, but do not completely close the door of lunar composition. In addition, our result and other dynamic models[71,77,78] suggest that Kamoʻoalewa could originate from the $\nu_6$ secular resonance or the Flora family, while our spectroscopic analysis also indicates that Kamoʻoalewa has a similar composition to 7 objects in the Flora family. This at least indicates that the scenario we proposed in this study is also reasonable and should be seriously considered.

**Regolith state of Kamoʻoalewa**
It was previously believed that small, fast-spinning asteroids are less likely to retain regolith due to the lower gravity and high centrifugal forces. However, recent calculations of two small, rapidly rotating NEAs, (499998) 2011 PT[83] (~35 m diameter, ~10 min rotation period) and 2016 GE1[84] (~34 s rotation period), show very low values for their thermal conductivity and thermal inertia, implying the presence of a fine regolith. Additionally, the latest thermal inertia calculation based on observations of the Yarkovsky effect suggested that Kamoʻoalewa has a low thermal inertia value of $155^{+90}_{-45}$ J m$^{-2}$ K$^{-1}$ s$^{-1/2}$ or $181^{+95}_{-60}$ J m$^{-2}$ K$^{-1}$ s$^{-1/2}$, indicating that its surface is covered by a regolith layer with grain size < 3 mm[53]. Our SW simulation experiments for slab (Fig. 2) and powder (Fig. 3, Supplementary Fig. 2) of LL chondrite also support a fine regolith rather than coarse material on the surface of Kamoʻoalewa. This is consistent with the calculation that Kamoʻoalewa could significantly retain size < 1 cm regolith using the gravity-adhesion-centrifugal force balance method and accounting for the competitive effect between thermal fatigue-caused regolith generation and impact-driven particle loss[54]. These allow the Tianwen-2 probe to conduct regolith sampling in the future.

Additionally, as seen in 40 mJ × 80 times irradiated LL chondrite powder (Fig. 4), we predict that amorphous rims and abundant npFe$^0$ particles will exist in the regolith of Kamoʻoalewa. npFe$^0$ particles are the main SW products that darken and redden asteroid surfaces and have also been found in Itokawa returned grains[20,26] and in laser-irradiated olivine[41,59,85].

**Surface evolution mechanisms of Kamoʻoalewa**
SW and resurfacing may have jointly contributed to Kamoʻoalewa's ultra-highly space-weathered surface. Since we simulated SW using loose powder for the first time, which is closer to the regolith state of asteroids than previously used compacted powders[41] or slabs, we developed a new model to convert irradiation parameters to the SW timescale of NEAs. We first tested the validity of this model using Itokawa and determined a SW timescale of $2.94 \times 10^6$ yr for Itokawa (Methods). This is close to the exposure age of return particles estimated from cosmic-ray-produced $^{21}$Ne content ($3 \times 10^6$ yr[26]), Brunetto's SW spectral model[86] ($2 \times 10^6$ yr[87]), and crater frequency statistics ($3.2 \times 10^6$ yr[88]). Further, the model shows that 40 mJ × 80 times of irradiation in Fig. 4 is equivalent to $9.33 \times 10^6$ yr of SW at 1 AU (Methods). The SW timescale of Kamoʻoalewa in near-Earth space, therefore, is estimated to be about $9.33 \times 10^6$ yr, 3.17 times that of Itokawa. For 1-m to 1-km LL chondrite-compositional NEAs originating from the Flora family, the dynamical lifetimes are between $3.45–11.93 \times 10^6$ yr[8]. This dynamical lifetime is comparable to the SW timescale of Kamoʻoalewa, probably allowing Kamoʻoalewa to evolve such an extremely red spectrum in near-Earth space.

Asteroid resurfacing mechanisms have been summarized as planetary encounters, collisions (and small impacts), Yarkovsky-O'Keefe-Radzievskii-Paddack effect (YORP) spin-up, and thermal fatigue[25]. First, Kamoʻoalewa's current minimum orbital intersection distance with Earth is 0.034 AU, which exceeds the range of Earth encounters (5–16 times the Earth's radius[25], 0.002–0.007 AU). Additionally, quasi-satellites generally do not experience close flybys with Earth[32]. Therefore, Earth encounter is implausible to refresh Kamoʻoalewa's regolith. Second, assuming a 1 cm meteoroid can significantly refresh Kamoʻoalewa's surface, such an impact event occurs per $2 \times 10^7$ yr (Methods). This is longer than the SW timescale and potential dynamical lifetime of Kamoʻoalewa. That is, impact-driven surface renewal is unlikely to have removed Kamoʻoalewa's regolith. Third, calculations for Kamoʻoalewa suggested that the current rotation state can result in the loss of centimeter-sized grains but not smaller ones or dust[54]. Our calculation shows that it only takes $4.27–42.7 \times 10^4$ yr at 1 AU for Kamoʻoalewa to accelerate to its current rotation state (Methods), 1–2 orders of magnitude smaller than the SW timescale. This indicates that large grains were lost very early, while small particles have been accumulating since they were produced. During asteroid regolith evolution, larger grains are fresher, whereas smaller particles generally are space-weathered[22]. Consequently, the long-term loss of large (fresh) grains and the accumulation of small (weathered) particles may have accelerated Kamoʻoalewa's surface reddening. Fourth, early experiments and calculations suggested that[80] thermal fatigue destroys rocks more quickly than micrometeoroid impacts on asteroids, but the former does not lead to regolith losses at perihelion > 0.3 AU (perihelion of Kamoʻoalewa is 0.898 AU). This indicates that the regolith at Kamoʻoalewa has a stable generation mechanism, and as discussed above, the surface of Kamoʻoalewa is covered by regolith. Thermal fatigue refreshes asteroid surfaces by breaking up the regolith grains and exposing fresh interior faces. At 1 AU, exposing the inner fresh faces of a 1 cm grain by thermal fatigue requires longer than $\sim 2 \times 10^7$ yr[80], whereas fast

SW reddening needs only about $10^6$ yr[89]. This indicates that SW is dominant in the competition between SW and resurfacing.

Notably, despite the possible evolution mechanisms discussed above, they are restricted to the near-Earth region. If the situation in which Kamoʻoalewa was once in the main belt is also considered, it may have lost large particles earlier and required a shorter SW timescale in near-Earth space to produce such a spectrum. However, it is currently unclear when Kamoʻoalewa was ejected from the Flora family, how long it was exposed in the main belt space, and what path it took to migrate to the near-Earth region. Therefore, the mechanism by which Kamoʻoalewa generates ultra-high SW is still an open question worth discussing later, which is expected to be finally answered by the future Tianwen-2 mission.

**Significance of the Tianwen-2 mission**

To date, Hayabusa, Hayabusa2, and OSIRIS-REx have all targeted slowly rotating rubble-pile asteroids larger than 100 meters in size for sample return. Limited by spatial resolution, ground-based telescopic observations also focus more on asteroids with sizes > 100 m. Although smaller objects are more numerous and pose a greater impact threat to the Earth, we know little about their surface characteristics and evolution mechanism. Kamoʻoalewa, as a sub-hundred-meter NEA[33], is expected to provide key clues to this issue. For example, why doesn't Kamoʻoalewa follow the observed rule for large Q/Sq/S asteroids[1]: the smaller the perihelion, the younger the surface? Additionally, Kamoʻoalewa and two small and rapidly rotating NEAs similar to Kamoʻoalewa, namely (499998) 2011 PT[83] and 2016 GE1[84], were calculated to show extremely low thermal conductivity (and thermal inertia) and potentially fine regolith. Therefore, two questions will hopefully be answered in the future: (1) Do all small, rapidly rotating, silicate-rich NEAs exhibit an extremely red spectral slope and develop a fine regolith surface? (2) If so, how do SW and refreshing processes together affect their surfaces? Finally, the returned sample will provide a definitive answer to the nature of Kamoʻoalewa: a piece of the Moon or an LL-chondrite-like body?

The Tianwen-2 probe will also orbit and monitor an active asteroid, 311P/PANSTARRS, which belongs to the Flora family and is losing particles[90], making it a potential main belt Q-type body with a fresh surface. If so, Kamoʻoalewa and 311P (as SW-strong and SW-fresh endmembers, respectively) are expected to compare well with Itokawa (SW-intermediate). This will enhance our understanding of the SW differences between the near-Earth region and the main belt space[80], especially the SW rate, as it is currently still highly controversial[91] among different models derived from Earth-ground observations[89,92,81,93] and simulation experiments[41,86,94]. Meanwhile, future sample analyses for Kamoʻoalewa are expected to provide key clues to the connection between Earth's LL chondrite falls and Kamoʻoalewa-like small NEAs[7].

## Methods

**Spectrally link Kamoʻoalewa to LL chondrite**

Here, we add some descriptions to supplement how we use reflectance spectra to identify analogues of Kamoʻoalewa. First, due to the Band I center wavelength position being diagnostic, 1.001±0.028 μm suggests that Kamoʻoalewa highly resembles LL chondrites in composition, and is less similar to L-chondrites and andesitic achondrite, rather than olivine, brachinites (mainly composed of olivine), H-chondrites, five primitive achondrites (brachinites, acapulcoites, lodranites, winonaites, and ureilites), and basaltic achondrites (Fig. 1b). Second, due to iron-nickel metal in metal-silicate systems being able to contribute an extremely red slope[49,95], we also consider the possibility of pallasites (composed mainly of olivine and metal) and mesosiderites (composed mainly of pyroxene and metal). However, pallasites show an absorption center at a longer wavelength (1.045–1.095 µm), and mesosiderites show an absorption center at a shorter wavelength (0.915–0.93 µm) (Fig. 1b), inconsistent with the 1.001±0.028 µm of Kamoʻoalewa. Even with SW, simulation experiments on olivine[40–42] and $Fe^{2+}$-bearing silicate meteorites[43,44] have suggested that the Band I center does not generally shift more than ±10 nm. This makes fresh silicate materials that did not show a match to the Band I center of Kamoʻoalewa (listed above) still fail to match Kamoʻoalewa after SW. Third, our laser irradiation experiments on H and L chondrites also show that neither the fresh nor space-weathered H and L chondrites match Kamoʻoalewa (Fig. 5). Fifth, the andesitic achondrite is mainly composed of pyroxene, feldspar, and quartz, and that its spectrum is similar to HED meteorites (daughters of V-type asteroids, rich in pyroxene and extremely insensitive to space weathering), it is less likely to contribute a Kamoʻoalewa-like red spectrum than the H and L chondrites. Sixth, 1.001±0.028 µm also excludes enstatite achondrites, which are generally featureless, although a few samples show a weak absorption at 0.89 µm[44]. It also rules out enstatite chondrites, which are generally featureless[49]. The iron meteorites are also ruled out because they are featureless[49]. Seventh, Kamoʻoalewa shows a 0.45–2.194 µm spectral slope as $0.723^{+0.076}_{-0.071}$ µm$^{-1}$. Such a high spectral slope could be seen in the carbonaceous chondrite Tagish Lake[96] and D-type asteroids (Fig. 7), but they lack an obvious Band I absorption[96]. Even if there is absorption, the band depth is usually shallow (< 3%). The carbonaceous composition is therefore ruled out for Kamoʻoalewa. Eighth, as described in the main text, we found that Kamoʻoalewa does not match the spectral parameters or space weathering patterns of the three celestial materials, but laser-irradiated LL chondrite powder does, suggesting that laser-irradiated LL chondrite powder is more credible for matching Kamoʻoalewa's composition than the three lunar materials. Regarding a comparison of the spectra of Kamoʻoalewa, the lunar Giordano Bruno crater, and other asteroids, please refer to the main text.

**Dynamical source region calculation**
Kamoʻoalewa's source region was estimated by extracting the likelihood that it originated in a specific source region or entered the near-Earth object (NEO) region through a specific escape route from an evolutionary model of the NEO population[10] based on its current semimajor axis $a$, eccentricity $e$, inclination $i$, and absolute magnitude $H$ (data are from the JPL web). The evolutionary model of the NEO population considers 3 possible source regions (Hungaria group, Phocaea group, and Jupiter-family comets) and 4 possible escape routes from the main asteroid belt (the $v_6$ secular resonance and nearby mean-motion resonances, the 3:1 mean-motion resonance with Jupiter and nearby lesser resonances, the 5:2 mean-motion resonance with Jupiter and nearby lesser resonances, and the 2:1 mean-motion resonance with Jupiter and nearby lesser resonances). The probability of origin for an NEO is obtained by extracting the contribution from each of these 7 source regions or escape routes in the *(a, e, i, H)* cell containing the NEO in question.

The METEODOM model[8] contains 52 asteroid families as source regions and requires presetting the composition of a small object. We input the *a*, *e*, and *i* of Kamoʻoalewa and selected the "All LL" option in the online version of the model to obtain the probabilities.

**Space weathering simulation experiments**

To determine if the extremely red spectral slope of Kamo'oalewa can be contributed by the SW process, we conducted laser irradiation experiments on a fresh LL5/6 chondrite, Kheneg Ljouâd. A slab and a loose powder with a grain size < 45 μm were irradiated here. Using the precision balance to weigh the mass of a 10 mm × 10 mm × 3 mm slab, we obtained its bulk density as 3357 kg m$^{-3}$. We also placed the loose powder into a 10 × 10 × 3 mm sample cell and weighed the mass, obtaining the bulk density of the powder sample as 760 kg m$^{-3}$. Furthermore, using a laser particle size analyzer, we analyzed the powder's particle size distribution and obtained its mean size of 29.17 μm. We also measured the mineralogical composition of slab samples using TIMA (TESCAN Integrated Mineral Analyzer) located at the Nanjing Hongchuang Geological Exploration Technology Service Co., Ltd. The result shows that Kheneg Ljouâd is composed of 57.73 vol.% olivine, 19.87 vol.% orthopyroxene, 5.61 vol.% diopside, 12.23 vol.% plagioclase, 2.62 vol.% troilite, 0.05 vol.% nickel-bearing troilite, 0.56vol.% tetrataenite, 0.76 vol.% chromite, 0.03 vol.% ilmenite, 0.12 vol.% orthoclase, and 0.38 vol.% apatite. We did not analyze the geochemical composition of Kheneg Ljouâd, however, the measurement results from the Meteoritical Bulletin Website can serve as a supplement: olivine $Fa_{31.0\pm0.2}$, low Ca pyroxene $Fs_{25.0\pm0.4}Wo_{2.1\pm0.2}$, high Ca pyroxene $Fs_{10.7}Wo_{43.1}$ and $Fs_{11.0}Wo_{43.0}$, feldspar $Ab_{84.4\pm2.2}An_{10.6\pm0.3}Or_{5.0\pm2.3}$, and tetrataenite (at%) $Fe_{42.9\pm0.2}Co_{2.1\pm0.1}Ni_{54.9\pm0.2}$. Then, we measured the reflectance spectra and performed laser irradiation experiments.

In the first step, we measured the reflectance spectra of fresh Kheneg Ljouâd slab and powder over the 0.35–2.5 μm wavelength range using the ASD FieldSpec 4 Hi-Res spectrometer at the Institute of Geochemistry, Chinese Academy of Sciences. This spectrometer is equipped with a 512-element silicon array detector (0.35–1.0 μm) and two Graded Index InGaAs detectors (1.001–1.8 μm and 1.801–2.5 μm) for collecting VIS-NIR reflectance spectral signals across wavelengths from 0.35 to 2.5 μm. The light source is an ASD illuminator featuring a 57 W quartz-tungsten-halogen lamp, capable of providing stable illumination across a range of 0.35 to 2.5 μm for spectral measurement. The spectral resolution is 3 nm at 0.7 μm, and 8 nm at both 1.4 μm and 2.1 μm. The spectral sampling is 1.4 nm in the 0.35–1.0 μm range and 1.1 nm in the 1.001–2.5 μm range. The measurement was conducted with an incidence angle of 30°, an emission angle of 0°, and a phase angle of 30°. A Spectralon Diffuse Reflectance Standard white plate was used during spectral measurements. We also measured dark current with the same set-up and subtracted it from both the sample and standard spectra prior to division of sample by standard.

In the second step, we conducted a laser irradiation experiment on a fresh slab and powder to simulate the spectrum contributed by micrometeoroid bombardments. To avoid the powder being too thick so that the laser cannot irradiate the lower layer of powder, based on past experience, we only loaded 0.3 g of powder sample in a 40 mm × 40 mm × 40 mm colorless glass box. At this time, the sample volume is 40 mm × 40 mm × ~0.27 mm. Here, the sample volume is estimated by dividing the mass (0.3 g) by the density (760 kg m$^{-3}$). Thereby, the powder particles can be irradiated sufficiently. The experiments were conducted by the 1064 nm Continuum Minilite II nanosecond pulse laser system (pulse width was 6 ns) located at the China University of Geosciences. The spot diameter of the laser irradiated on the sample is 0.5 mm, and the vacuum pressure is $1.4 \times 10^{-3}$ Pa. We set the laser frequency to 10 Hz, which means that 10 pulses are shot continuously at each point (a circular area with a diameter of 0.5 mm) per second. The laser energy was set to 40 mJ per pulse. Therefore, theoretically, each point receives 0.4 J of laser energy per second. The laser then irradiated the next point until the entire 40 mm × 40 mm area was scanned. Each scan took 30 minutes. For the slab sample, we scanned 4 times, 8 times, and 12 times, so the laser parameters were 40 mJ × 40 times, 40 mJ × 80 times, and

40 mJ × 120 times, respectively. For the powder sample, we scanned 1 time, 2 times, 4 times, and 8 times, so the laser parameters were 40 mJ × 10 times, 40 mJ × 20 times, 40 mJ × 40 times, and 40 mJ × 80 times, respectively. Finally, we measured the reflectance spectra of the laser-irradiated slab and powder under the same conditions as the fresh samples.

Further, using the FEI Scios Dual Beam scanning electron microscope located at the Institute of Geochemistry, Chinese Academy of Sciences, we observed the microscopic morphology of LL chondrite powder after 40 mJ × 80 times irradiation. Then, using the FEI Talos F200X field emission transmission electron microscope located at the Sinoma Institute of Materials Research, we observed the SW micro characteristics. The results are shown in Fig. 4.

Using the same method as for the LL chondrite powder, we also irradiated H6 (Xingyang) and L6 (Beni M'hira) chondrite powders with a size of < 45 μm at 40 mJ × 20, 40, 60, and 80 times, respectively. We then measured and analyzed their fresh and irradiated spectra. The results are shown in Fig. 5. Additionally, note that some of the spectra we measured for H, L, and LL chondrites present small absorption features near 2.1 μm. This is due to incomplete correction of the reference white plate and does not affect the analysis of the spectral results.

**SW timescale estimation**

We used formulas (1) to (4) to estimate the simulated SW timescales for NEAs, which are modified from the previous methods for dense samples (namely, slab[44] and compacted powder[41]). This method assumes that all kinetic energy generated by micrometeoroid bombardment of the asteroid surface is converted into heat and that the micrometeoroid flux is steady. At 1 AU, the impact flux ($F$) of a dust particle with a diameter ~$10^{-6}$ m (mass $m = 10^{-15}$ kg) is about $10^{-4}$ m$^{-2}$ s$^{-1}$ (ref. 97). Assuming that the average speed ($v$) of each dust particle hitting the asteroid is $2 \times 10^{4}$ m s$^{-1}$, then within one year (time $T = 365 \times 24 \times 3600$ s), according to formula (1), the energy deposited per square meter per year on the asteroid surface, $A$, will be $6.307 \times 10^{-4}$ J.

$$A = \frac{1}{2}mv^2 FT \tag{1}$$

The laser energy density ($B$) received by powder particles can be calculated by formula (2):

$$B = \frac{\sum_i (0.4 \times 1800 \times 0.1 \times 0.33 \times S_i \times T_i)}{4 N_p \pi r^2} \tag{2}$$

where 0.4 is, ideally, the total energy deposited at each point (circular area with a diameter of 0.5 mm) within 1 s of each scan, in J. 1800 is the total time of each scan, in s. Since we irradiated loose powders rather than compacted powders, we noticed that when the first pulse shot the sample, the laser pushed the sample away, forming a circular area with a diameter of approximately 2.5 mm without a sample (see Supplementary Fig. 3a). This will result in the 2nd to 10th pulse per second not irradiating the sample, so we introduce an irradiation factor of 0.1. The laser's displacement effect also results in the absence of samples at the next 2nd to 3rd irradiation points, therefore, we introduce the first area factor of 0.33. The laser's displacement effect also causes approximately 30% of the area to be free of sample for each scan after the first one (see Supplementary Fig. 3b), so, we introduce the second area factor, $S_i$. $S_i = 1$ when $i = 1$, and $S_i = 0.7$ when $i > 1$. $i$ is the number of scans. Since the laser also evaporates some of material onto the quartz window, resulting in laser energy loss before reaching the sample, we also introduce a transmittance factor, $T_i$. We use the energy meter to measure that $T_i = 0.83$ after the 1st scan, and $T_i = 0.65$ after the 8th scan. That is, $T_i$ decreases by about 0.025 each scan as $i$ increases.

We assumed that each particle could be irradiated and that the shape of the particle was a sphere with a radius of $r$. Thus, the numerator in formula (2) represented the total surface area of all particles. In this study, $r = 1.4585 \times 10^{-5}$ m. $N_p$ is the number of particles loaded in each experiment, which can be calculated by formula (3):

$$N_p = \frac{0.3}{\frac{4}{3}\pi r^3 \rho} \tag{3}$$

where 0.3 is the powder weight loaded in the experiment, in g. $\rho = 3357$ kg m$^{-3}$ is the density of each particle.

Thereby, the simulated SW timescale ($C$, in yr) can be calculated by formula (4):

$$C = \frac{B}{A} \tag{4}$$

To evaluate the validity of this method, we first used it to estimate the SW timescale of Itokawa. We found that when scanning 2 times ($i = 2$), the produced spectrum is close to that of Itokawa (Supplementary Fig. 4). At this time, the SW time scale $C = 2.94 \times 10^6$ yr. This timescale is close to the exposure age of Itokawa's return particles estimated by cosmic-ray-produced $^{21}$Ne content ($3 \times 10^6$ yr[26]), Brunetto's spectral model[86] ($2 \times 10^6$ yr[87]), and surface exposure age calculated from impact crater frequency ($3.2 \times 10^6$ yr[88]), indicating that this method is applicable to LL chondrite-compositional NEAs with loose regolith. Then, for 40 mJ × 80 times experiment ($i = 8$), our calculated SW timescale $C = 9.33 \times 10^6$ yr.

**Impact-driven resurfacing timescale estimation**
We used formulas (5) to (8) to estimate the impact-driven resurfacing timescale on Kamo'oalewa. This method is based on the meteoroid-Earth impact flux model[98] and accounts for the surface area difference between the Earth and Kamo'oalewa. The cumulative number ($N$) of meteoroids colliding with the Earth per year could be described as formula (5)[98]:

$$\log N = 1.568 - 2.7 \log D \tag{5}$$

where $D$ is the diameter of the meteoroid, in m. Assuming Kamo'oalewa has the same impact density as Earth, the cumulative number of meteoroids colliding with the Kamo'oalewa per year ($N_{asd}$) could be described as formula (6):

$$N_{asd} = \frac{4\pi R^2}{4\pi (6.371 \times 10^3)^2} N \tag{6}$$

where $R$ is the radius of Kamo'oalewa, in m. Thereby, the impact-driven resurfacing timescale ($T_{asd}$) was calculated as formula (7):

$$T_{asd} = \frac{1}{N_{asd}} \tag{7}$$

In formula (6), $R$ is determined by the empirical formula (8):

$$1000R = \frac{1329}{2\sqrt{Pv}} 10^{-0.2H} \tag{8}$$

where $Pv$ is the visual geometric albedo, and $H$ is the absolute magnitude. For Kamo'oalewa, we used the $H = 24.33$ mag[33] and the $Pv = 0.1$. $Pv = 0.1$ is used here because the reflectance at 0.55 μm (this value of meteorites is close to the visible geometric albedo of the asteroid) of 40 mJ × 80 times laser irradiated Kheneg Ljouâd powder is 0.085 (Fig. 3), close to 0.1. Thereby, assuming a 1 cm diameter

($D$ = 0.01 m) meteoroid impacted the surface of Kamoʻoalewa and caused significant resurfacing, the $T_{asd}$ is calculated as $2 \times 10^7$ yr.

**YORP spin-up timescale estimation**

We used formulas (9) to (11) to estimate Kamoʻoalewa's YORP spin-up timescale, namely the time elapsed from no spin acceleration to the current spin rate. To facilitate calculation, we assume that Kamoʻoalewa is a regular three-axis ellipsoid. The torque exerted upon the asteroid by the YORP effect can be described as formula (9)[99]:

$$T_{YORP} = \frac{\emptyset C_a R^3}{c a^2 \sqrt{1-e^2}} \quad (9)$$

where $\emptyset$ = 1361 W m$^{-2}$ is the solar constant at 1 AU. $C_a$ is the dimensionless YORP coefficient lying in the range of 0.002 to 0.02 for most asteroid shapes[100]. $R$ is the asteroid radius, calculated by formula (8). $c = 2.998 \times 10^8$ m s$^{-1}$ is the light speed. $a$ and $e$ are the semimajor axis and eccentricity, respectively.

The angular acceleration caused by the torque is described in equation (10)[101]:

$$\frac{d\omega}{dt} = \frac{T_{YORP}}{I} \quad (10)$$

where $\omega$ is the angular velocity, in rad s$^{-1}$. $t$ is time, in s. $I$ is the moment of inertia:

$$I = \frac{1}{5}\left(\frac{4}{3}\rho\pi\frac{l_a l_b l_c}{8}\right)\frac{l_a^2 + l_b^2}{4} \quad (11)$$

For Kamoʻoalewa, we used $\omega = 3.9 \times 10^{-3}$ rad s$^{-1}$ (rotation period = 27 min[33]), $C_a$ = 0.002 to 0.02, $\rho$ = 3357 kg m$^{-3}$, $a$ = 1, and $e$ = 0.1, $R$ = 28.6 m. The lengths of the three axes of Kamoʻoalewa are used as $l_a$ = 69.4 m, $l_b$ = 58.5 m, and $l_c$ = 51.8 m (ref. [55]). Thus, assuming Kamoʻoalewa was accelerated to the current rotation period with a uniform angular acceleration at 1 AU, the YORP spin-up timescale, $\Delta t$, will be 4.27–42.7 × 10$^4$ yr.

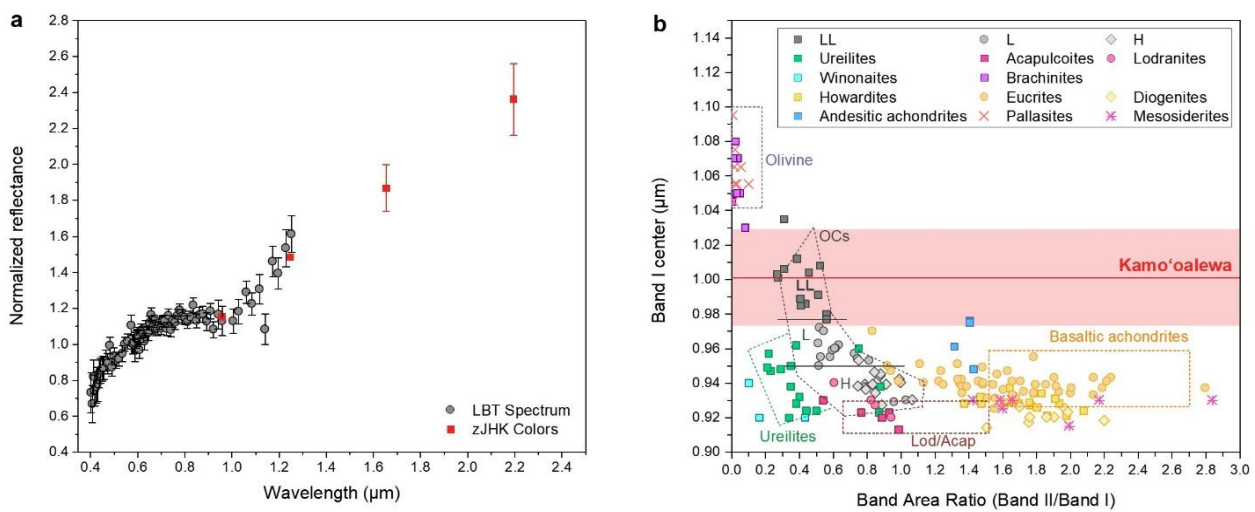

**Fig. 1 | Spectral matches of Kamo'oalewa to meteorites. a** Normalized (at 0.55 μm) VIS-NIR reflectance spectrum of Kamo'oalewa, which exhibits a Band I absorption feature. Kamo'oalewa's spectral data was previously obtained using the Multi-Object Double Spectrograph (0.4–0.95 μm) and the Utility Camera in the Infrared instrument (0.95–1.25 μm) on the LBT, along with broadband zJHK photometric color measurements[33]. **b** Comparison of band I center and band area ratio of Kamo'oalewa with meteorites. Using the Monte Carlo method to generate 10,000 spectra from the spectral data in **a** and removing the continuum, Kamo'oalewa's Band I center range, 1.001±0.028 μm (red line and red range, error is 1σ), highly overlaps with LL chondrites and partly overlaps with L chondrite and andesitic achondrite. A box containing primitive achondrites (lodranites, acapulcoites) is cited from ref. 102. The box for ureilites is cited from ref. 103. Boxes for olivine, basaltic achondrites, and boundary lines of OCs are cited from ref. 104. Background data of meteorites are described in Supplementary Data 1.

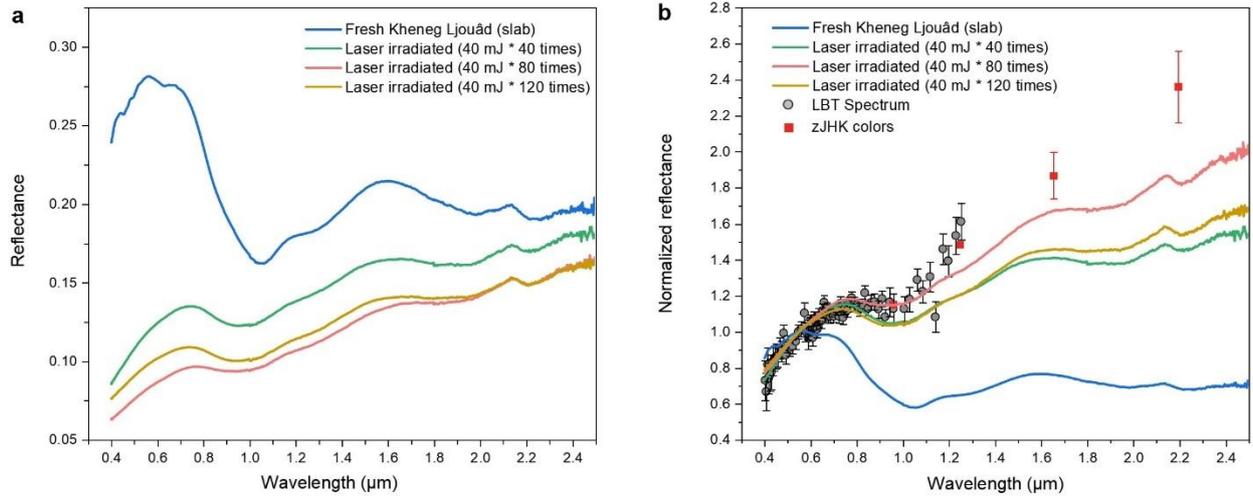

**Fig. 2 | Reflectance spectra of fresh and laser-irradiated LL chondrite slabs. a** Reflectance decreases as irradiation energy increases, but it starts to rise again once the energy exceeds 40 mJ × 80 times, indicating that the simulated SW reaches saturation at this energy (red line). **b** Spectral slope increases with the irradiation energy but starts to decrease when the energy exceeds 40 mJ × 80 times, meaning that the simulated SW reaches saturation at this energy (red line). However, even when the simulated SW is saturated, the corresponding spectral slope is still much flatter than that of Kamoʻoalewa. The spectra are normalized at 0.55 μm.

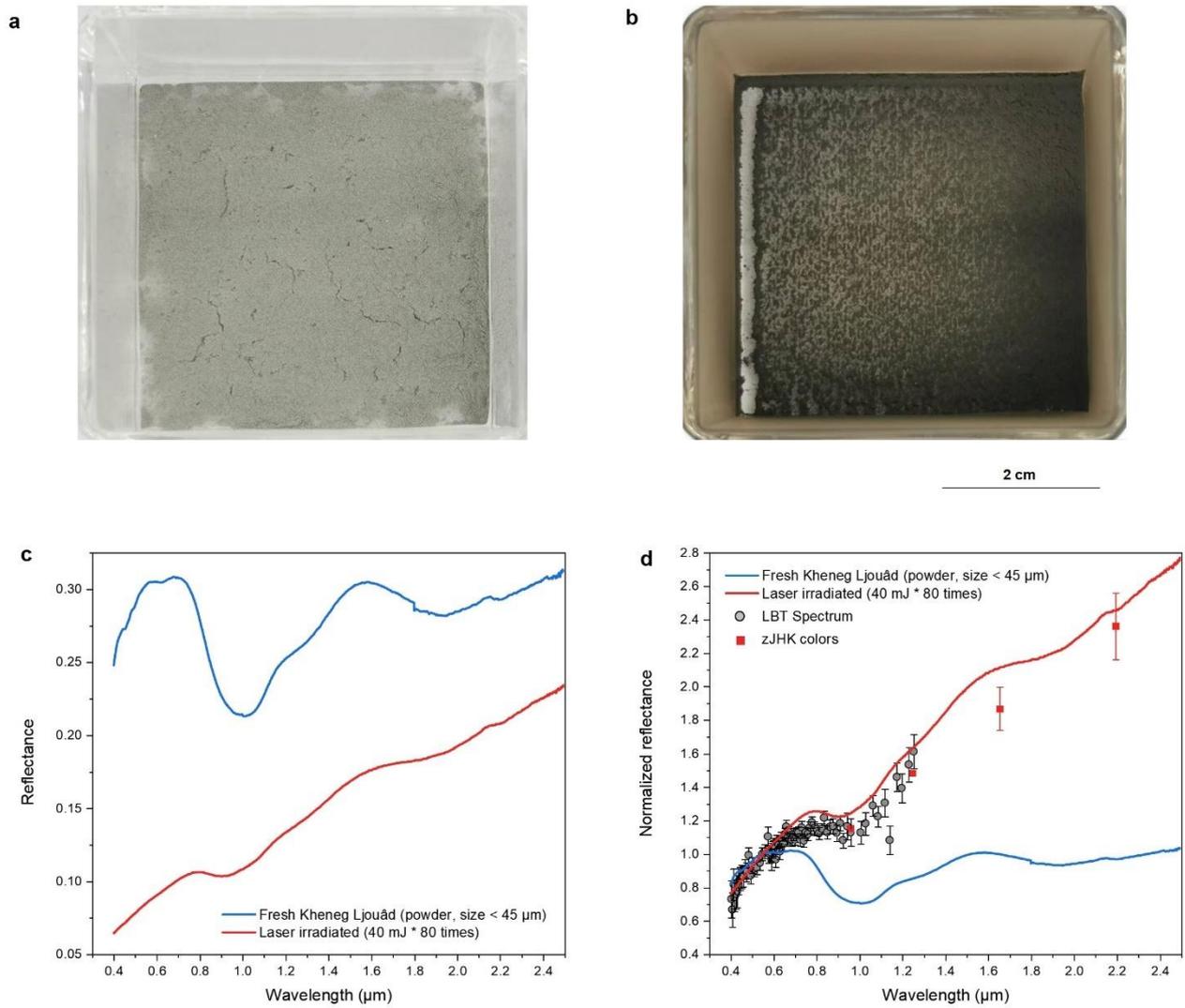

**Fig. 3 | Photos and reflectance spectra of fresh and laser-irradiated LL chondrite powder samples. a** Photo of fresh powder from the LL5/6 chondrite Kheneg Ljouâd, with a particle size of less than 45 μm. It displays a light gray color. **b** Photo of irradiated powder with 40 mJ × 80 times laser energy. It has a dark gray color. **c** Reflectance spectra of the powders in (**a**) and (**b**). After irradiation, there was a significant decrease in reflectance. **d** Normalized (at 0.55 μm) spectra of Kamoʻoalewa and the powders in (**a**) and (**b**). After irradiation, there was a significant increase in the spectral slope. The spectral curve of the irradiated powder closely matched that of Kamoʻoalewa.

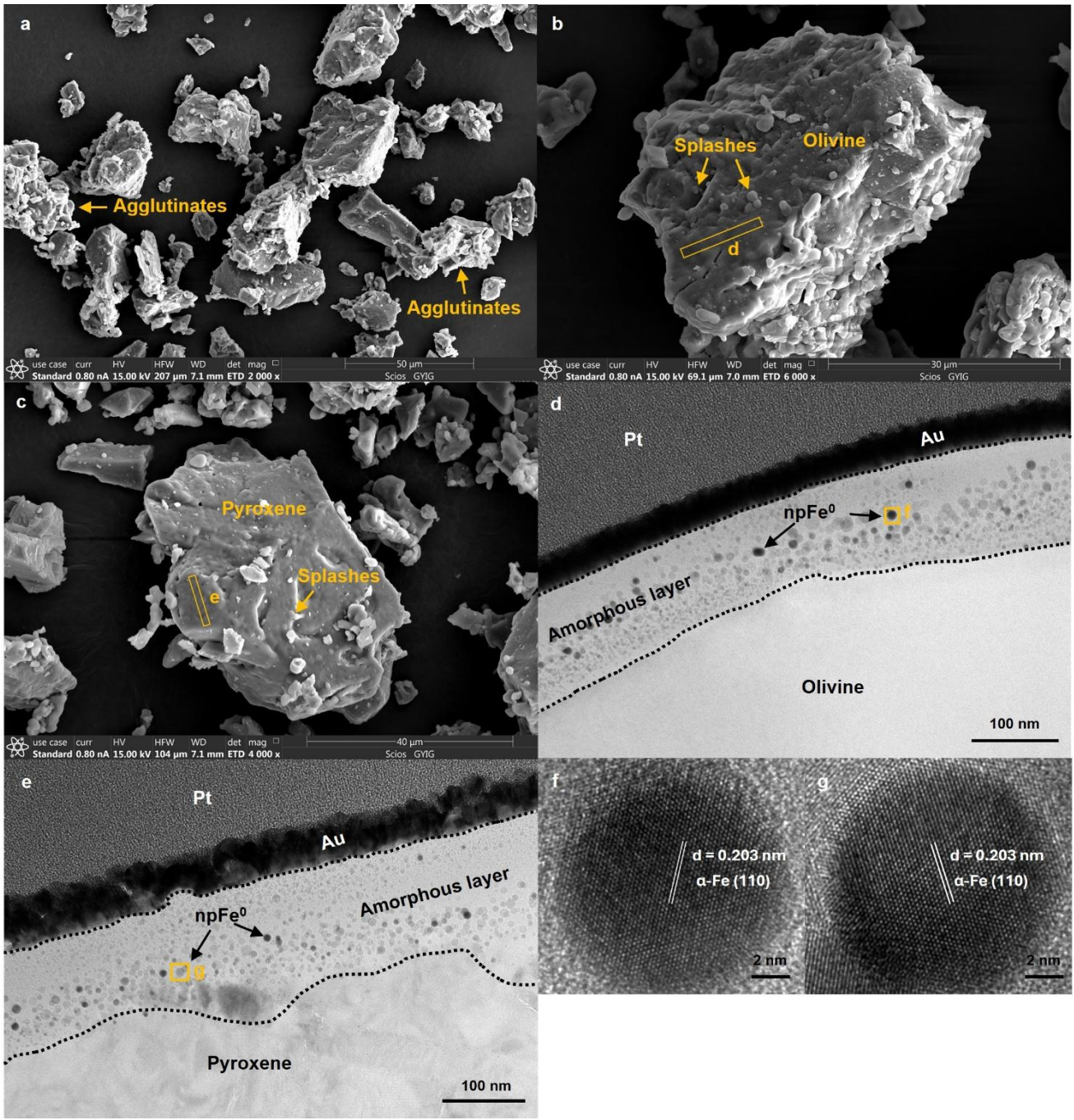

**Fig. 4 | Micro-characteristics of 40 mJ × 80 times laser-irradiated Kheneg Ljouâd powders observed by scanning electron microscopy (SEM) and transmission electron microscopy (TEM).**
**a** SEM secondary electron image of irradiated powder. **b** SEM secondary electron image of olivine grain in irradiated powder. **c** SEM secondary electron image of pyroxene grain in irradiated powder. **d** TEM bright field image of FIB section sampled from box in (**b**). **e** TEM bright field image of FIB section sampled from box in (**c**). **f** TEM high-resolution image of npFe$^0$ in (**d**). **g** TEM high-resolution images of npFe$^0$ in (**e**). After irradiation, the roundness of grains increased, and agglutinates appeared (**a**). Meanwhile, abundant fine sediments and splashes are distributed on the surface of olivine and pyroxene grains (**b**–**c**). Laser irradiation also produces amorphous rims and npFe$^0$ particles on the surface of olivine and pyroxene grains (**d**–**e**). The interplanar spacing is d = 0.203 nm, consistent with the spacing of a crystal lattice plane (110) of α-Fe (**f**–**g**).

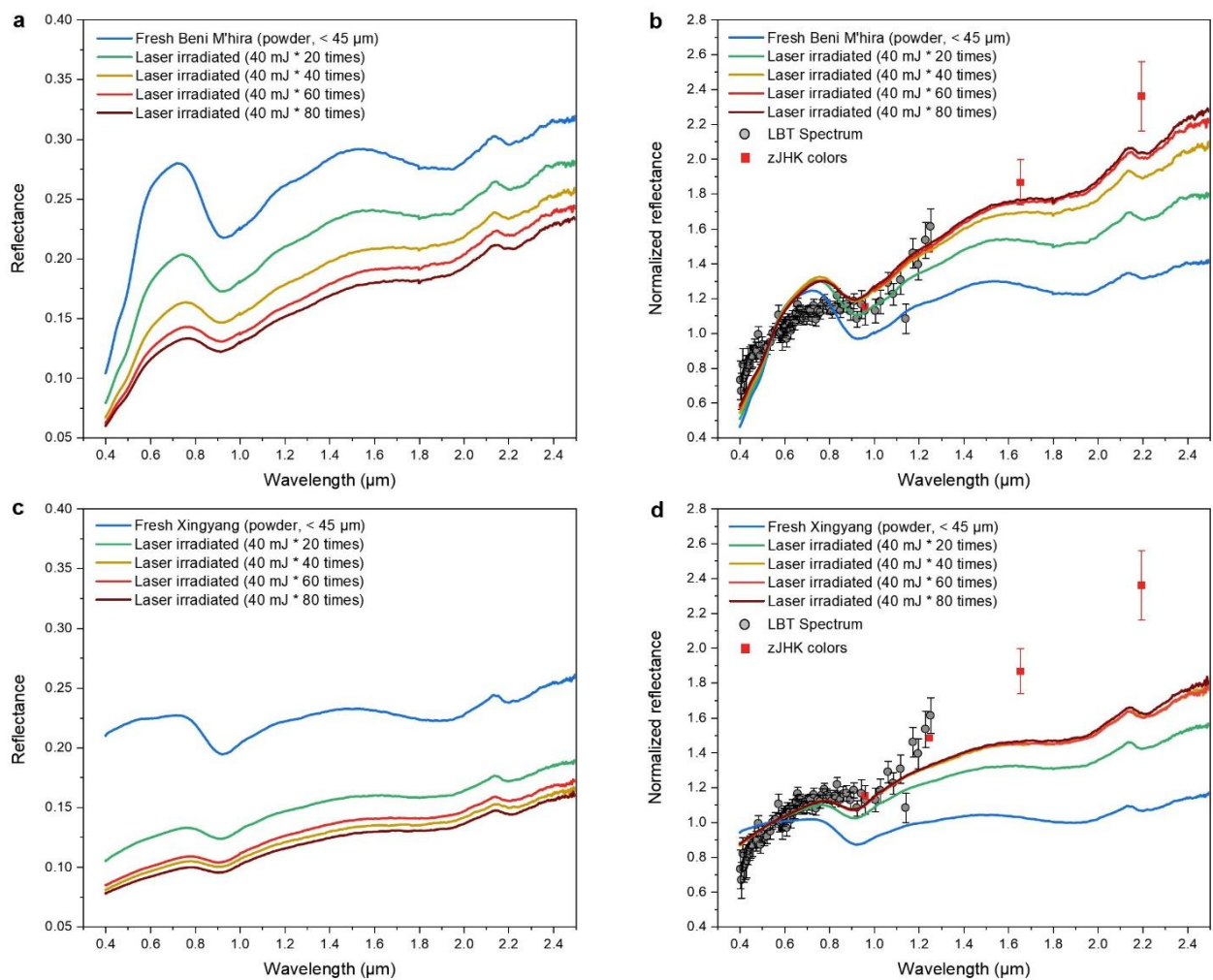

**Fig. 5 | Comparison of the reflectance spectra of Kamoʻoalewa, fresh, and laser-irradiated L and H chondrite powders (size <45 μm). a** Reflectance of Beni M'hira (L6 chondrite) decreases as the irradiation energy increases. **b** Spectral slope of Beni M'hira increases as the irradiation energy increases. **c** Reflectance of Xingyang (H6 chondrite) decreases as the irradiation energy increases. **d** Spectral slope of Xingyang increases as the irradiation energy increases. Neither fresh nor the laser-irradiated Beni M'hira and Xingyang match Kamoʻoalewa's spectral slope (**b** and **d**).

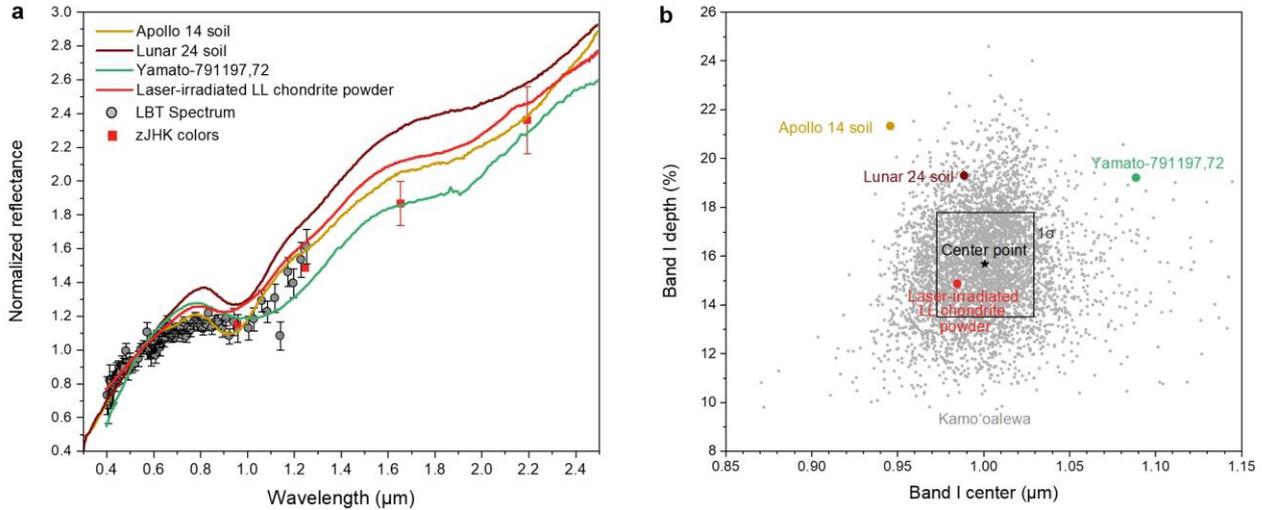

**Fig. 6 | Comparison of the reflectance spectra of Kamoʻoalewa, three lunar materials, and laser-irradiated LL chondrite (Kheneg Ljouâd) powder (size < 45 μm). a** Normalized reflectance at 0.55 μm of Kamoʻoalewa, three lunar materials, and 40 mJ × 80 times laser-irradiated Kheneg Ljouâd powder. Three lunar materials from the RELAB database, namely Apollo 14 soil (Spectrum ID C1LR90), Lunar 24 soil (Spectrum ID C1LR122), and lunar meteorite Yamato-791197,72 (Spectrum ID CALMCA), and laser-irradiated Kheneg Ljouâd powder, all show Band I absorptions and similar spectral slopes as Kamoʻoalewa. **b** Band I centers and Band I depths of Kamoʻoalewa, three lunar materials, and laser-irradiated Kheneg Ljouâd powder. Band I center and Band I depth of laser-irradiated Kheneg Ljouâd powder fall into the 1σ range of Kamoʻoalewa, while three lunar materials do not, suggesting that the former match Kamoʻoalewa better than the latter. The center point (asterisk symbol) means the average value of Kamoʻoalewa's data points.

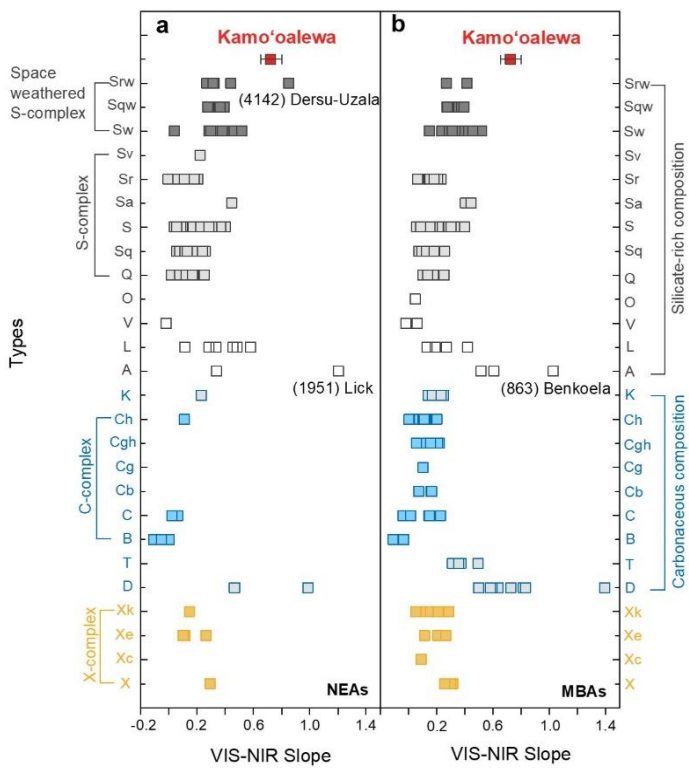

**Fig. 7 | Comparison of the VIS-NIR spectral slopes of Kamoʻoalewa with NEAs and MBAs. a** A-type (1951) Lick and Srw-type (4142) Dersu-Uzala show higher spectral slopes than Kamoʻoalewa. **b** A-type (863) Benkoela shows a higher spectral slope than Kamoʻoalewa. Asteroid spectral data and types are from MITHNEOS[1] and SMASS II[62], respectively. See Supplementary Data 1 for more details.

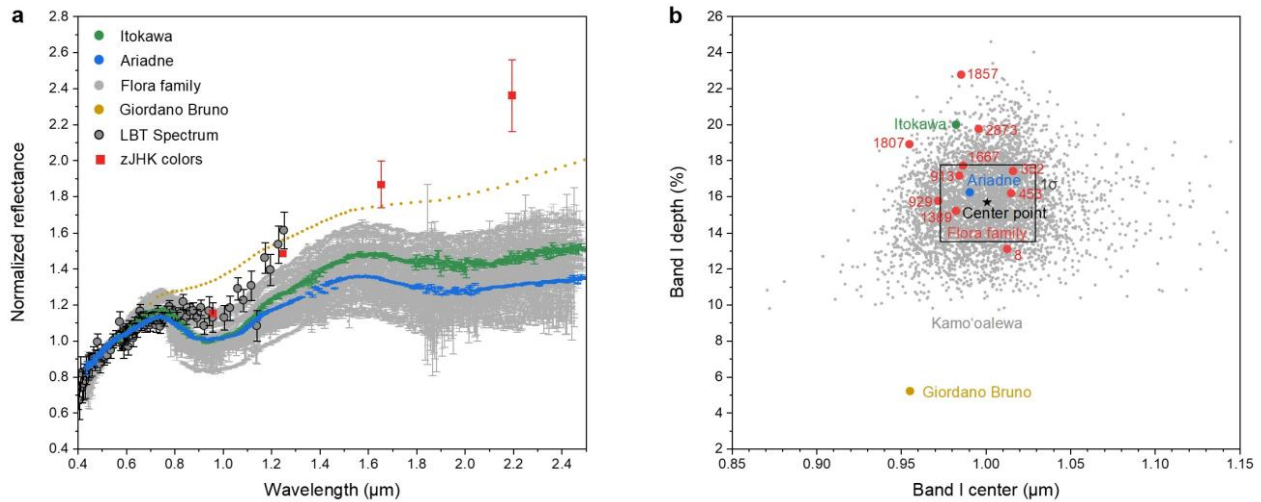

**Fig. 8 | Comparison of the reflectance spectra of Kamoʻoalewa, lunar crater Giordano Bruno, Flora family asteroids, and Itokawa. a** Normalized (at 0.55 μm) spectra of Kamoʻoalewa, lunar crater Giordano Bruno, 25 Flora family asteroids, and Itokawa. Kamoʻoalewa shows the reddest spectral slope. **b** Band I centers and Band I depths of Kamoʻoalewa, lunar crater Giordano Bruno, 11 Flora family asteroids, and Itokawa. Itokawa and 7 objects in the Flora family (43 Ariadne, 352 Gisela, 453 Tea, 913 Otila, 1667 Pels, 1857 Parchomenko, and 2873 Binzel) fall into the 1σ range of Kamoʻoalewa's Band I center and show deeper Band I depth than Kamoʻoalewa's average value (asterisk symbol). Giordano Bruno does not match the Band I center and the Band I depth of Kamoʻoalewa. The M³ spectral data for Giordano Bruno (M3G20090531T215442) is sourced from the PIPE web (Supplementary Data 1). Spectral data for Flora family asteroids and Itokawa are cited from ref. 12 and ref. 14, respectively. The center point (asterisk symbol) means the average value of Kamoʻoalewa's data points.